\documentclass[letterpaper]{article}
\usepackage{iccc}
\usepackage{graphicx}
\usepackage{natbib}
\usepackage{hyperref} 

\usepackage{times}
\usepackage{helvet}
\usepackage{courier}
\title{When happy accidents spark creativity:\\ Bringing collaborative speculation to life with generative AI}

\author{Ziv Epstein\textsuperscript{1*}, Hope Schroeder\textsuperscript{1*}, Dava Newman\textsuperscript{1,2} \\
\textsuperscript{1}MIT Media Lab\\
\textsuperscript{2}Human Systems Laboratory, Department of Aeronautics and Astronautics, MIT\\
\{zive, hopes, dnewman\}@mit.edu\\
\textsuperscript{*}These authors contributed equally
}
\begin{document} 
\maketitle
\begin{abstract}
\begin{quote}
Generative AI techniques like those that synthesize images from text (text-to-image models) offer new possibilities for creatively imagining new ideas. We investigate the capabilities of these models to help communities engage in conversations about their collective future. In particular, we design and deploy a facilitated experience where participants collaboratively speculate on utopias they want to see, and then produce AI-generated imagery from those speculations. In a series of in-depth user interviews, we invite participants to reflect on the generated images and refine their visions for the future. We synthesize findings with a bespoke community zine on the experience. We observe that participants often generated ideas for implementing their vision and drew new lateral considerations as a result of viewing the generated images. Critically, we find that the unexpected difference between the participant's imagined output and the generated image is what facilitated new insight for the participant. We hope our experimental model for co-creation, computational creativity, and community reflection inspires the use of generative models to help communities and organizations envision better futures. 
\end{quote}
\end{abstract}

\section{Introduction}
New methods in generative machine learning, such as GANs, have created an explosion of opportunities and possibilities for computational creativity. One possibility GANs afford is the empowerment of people in casual creation \citep{compton2015casual, epstein2020interpolating, berns2020bridging}. For such casual creation, text-to-image models built on technologies like CLIP \citep{radford2021learning} have enormous promise by providing people an intuitive and user-friendly possibility space to query GAN-generated images via prompts (see \citet{colton2021generative} for an overview). With such a technology, what are the possibilities for human creativity, and how might these applications impact communities? One opportunity for these text-to-image models is to aid in \textit{imaginative idea visualization} \citep{colton2021generative} as a way to bootstrap the creative process. An issue of particular relevance to this goal is the inherent variety/fidelity trade-off of generative models: as the outputs of models become more realistic, they become less diverse \citep{ramesh2022hierarchical}. Yet it remains unclear how this trade-off impacts downstream tasks like imaginative idea visualization. On one hand, high-fidelity outputs might provide helpful details for how a given idea might be actually implemented. On the other, ``happy accidents'' from diverse but low-fidelity outputs might help evolve an idea via lateral thinking. 

To explore how these two possibilities trade off, we design, deploy, and evaluate a computational creativity system for imagining and visualizing new ideas. Participants collaboratively speculated on utopian ideas for the future. These speculations were then fed as text prompts into a generative AI model to visually manifest them. We conducted a series of user interviews to learn about the experience from participants, and surface key themes. In this paper, we present a field report of the experience and use the system to trace broader questions about the social and collaborative aspects of creativity, such as when generative visual imagery can inspire ideas about the future, and how the variety/fidelity trade-off in generative models might impact creativity. Our work builds on previous work \citep{rafner2021utopian, rafner2020crea} which builds tools for people to express both hope and anxiety for the future via blending with generative models, and investigates  image features associated with dystopian versus utopian visions of the future.

This work builds on the tradition of speculative design and design fiction, which use design to imagine and prototype alternative worlds \citep{dunne2013speculative}. The social nature of the collaborative speculation allows individuals to build these alternative worlds together. Furthermore, it explores how organizations or communities could use speculative design and design fiction to chart a course for the future. 

There are several key contributions in this field report. First, we introduce a novel approach to prompt engineering, which leverages collaborative speculation from a facilitated co-design experience, rather than a single individual sourcing the prompts. Second, we highlight the importance of high-variance, low-fidelity images in inducing creative insights. Finally, we highlight the potential for computational creativity to aid in community dialogue and the collective elicitation of an organization's values. We explore this possibility by creating a zine to document and synthesize the findings, which we then present back to the community.

\section{Methods}
On October 8th, 2021, we organized a facilitated co-design experience at a solarpunk\footnote{Solarpunk is an artform and aesthetic that imagines near-distant futures where humans have become climate change-resilient and learned to live in harmony with nature.}-themed event at the MIT Media Lab, which had over 400 RSVPs. Participants in the exercise paired up and responded to the following prompt: ``How will we re-imagine the following categories in utopia? Team up with someone, and each pick one of the following sectors: (see Table ~\ref{table:categories}). Brainstorm how the future could be better at their intersection! Pick up a leaf and write your vision down on it. Add any visual representation you want. Tape the leaves to a stalk, add a flower, and put them in the solarpunk garden together!''

\begin{table}[h]
\begin{center}
\small
\begin{tabular}{ |c|c|c|c| } 
 \hline
 Money & Medicine & Cities & Transportation  \\ 
 Space & Agriculture & Music & Environment  \\ 
 Economy & Relationships & Family  & Healthcare  \\ 
 Arts &  Civil rights & Fashion  & Infrastructure  \\ 
 Trade & Social justice & Oceans  & Natural land \& Wildlife\\ 
 Education & Government & Energy  & Community  \\ 
 \hline
\end{tabular}
\end{center}
\caption{The 20 sectors used for prompt generation. Pairs of participants picked two to blend.}
\label{table:categories}
\end{table}

\begin{figure}[h]
\includegraphics[width=0.23\textwidth]{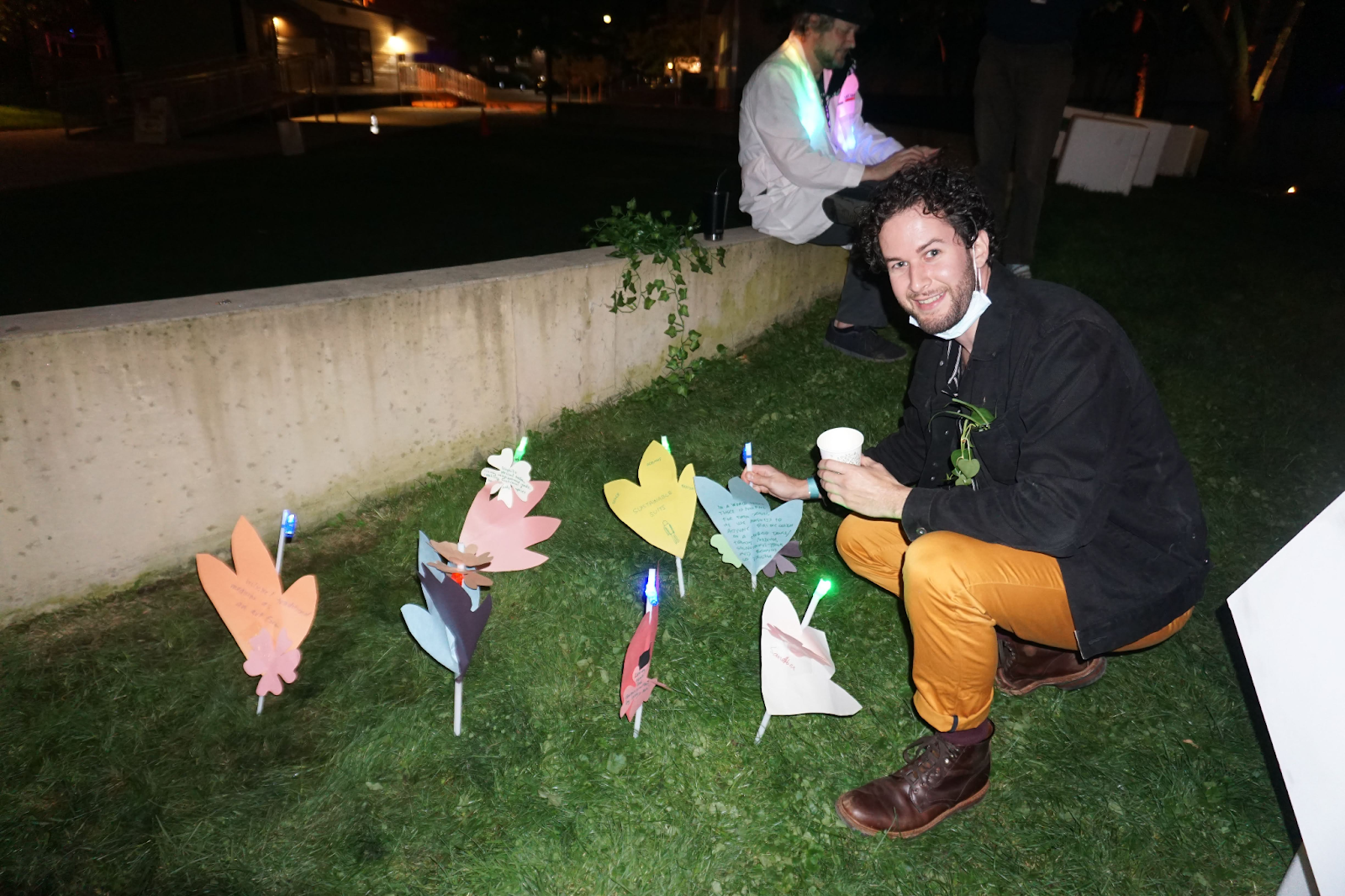}
\includegraphics[width=0.23\textwidth]{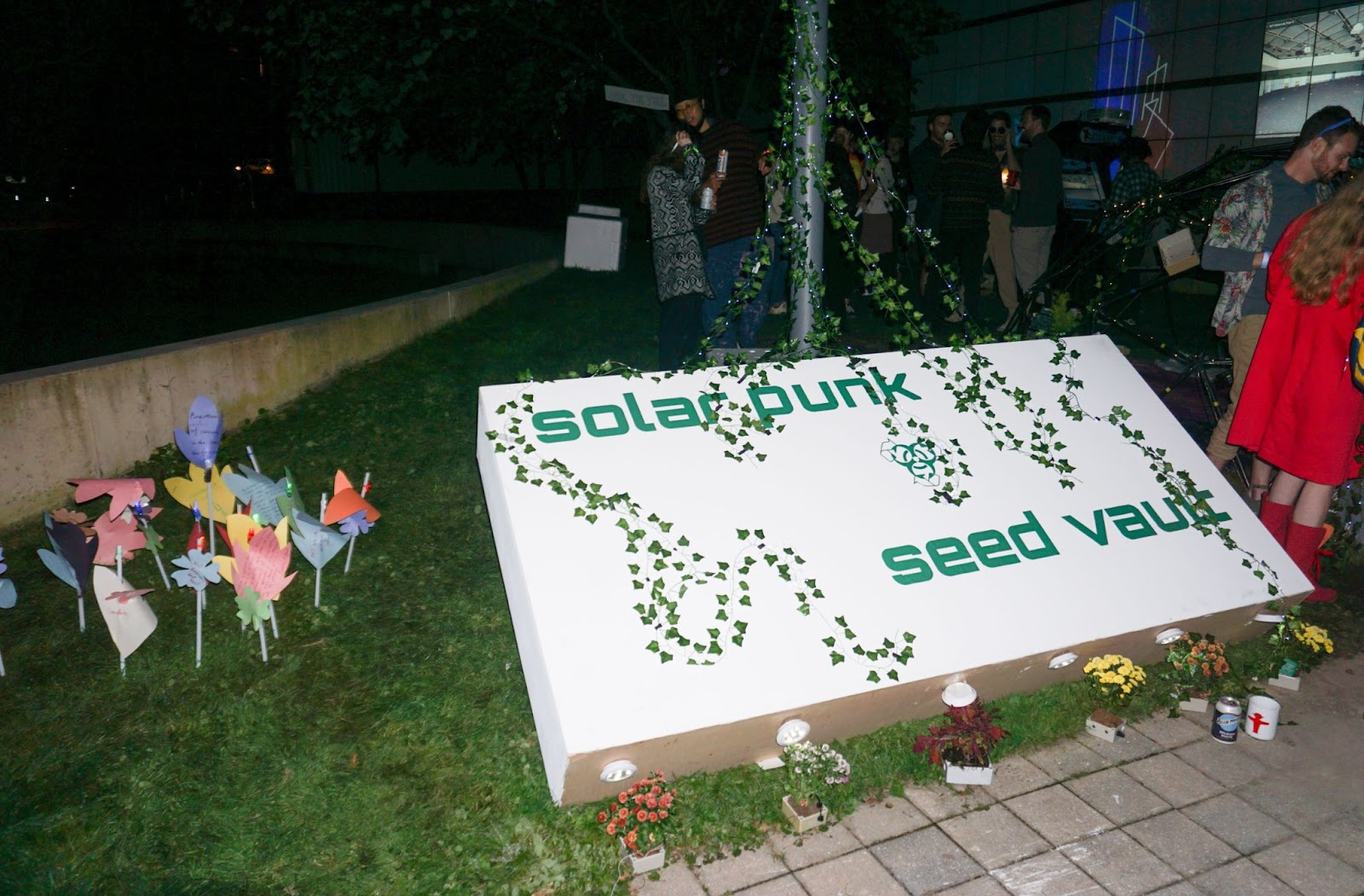}
\centering
\caption{Solarpunk garden where participants placed their written visions. The garden was conceptualized as a ``seed vault,'' a place for hopeful seeds for the future to be stored and preserved.}
\label{fig:solarpunk}
\end{figure}

After co-creating their visions, participants placed their flowers with the vision written on it in the solarpunk garden (Figure ~\ref{fig:solarpunk}). A total of 32 visions of the future were co-created by over 64 participants.

Next, we took the 32 visions, and ran them through VQGAN+CLIP, a common model of text-to-image synthesis. This produced visual representations of the participants' visions, and served as the output of the speculation experience. We then used these images for evaluation of the paradigm and incorporated them into a zine for additional community impact. 

\subsubsection{Evaluation via user interviews}

To evaluate the experience of facilitated speculation augmented with generative AI, we conducted a series of ten 15-20 minute semi-structured interviews from March 30th to April 11th, 2022 about 10 unique visions both in person and by video chat. 

First, to measure relatedness between prompt and image, we tested if participants could recognize the image generated by their prompt. For each participant, we paired the image that corresponded to their prompt with three additional images from other prompts and presented the set of four to the participant in random order. We invited participants to identify the image generated from the prompt they wrote in the original activity. Regardless of their answer, we then revealed the correct image. We then asked questions related to their interpretation of the image, starting by asking them to describe what they saw. Based on directions from this initial description, we then asked follow-up questions about if and how the image differed from their expectations. Finally, we asked if the participant had ideas for a follow-up prompt that better reflected their vision. After interviews were complete, we coded them using standard qualitative coding practices for themes that emerged across conversations \citep{saldana2021coding} .

\section{Results}
\begin{figure*}[h!]
\includegraphics[width=0.24\textwidth]{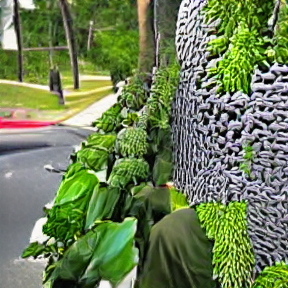}
\includegraphics[width=0.24\textwidth]{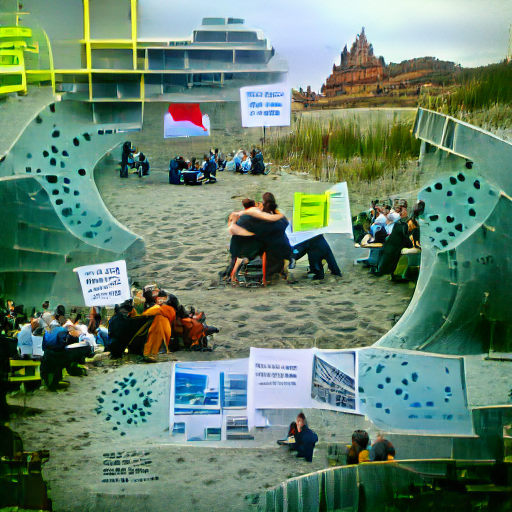}
\includegraphics[width=0.24\textwidth]{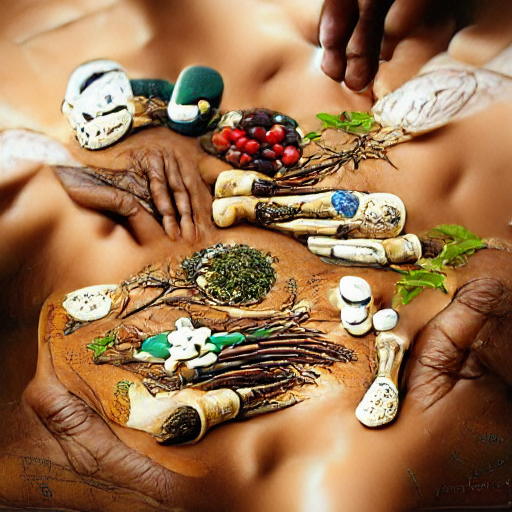}
\includegraphics[width=0.24\textwidth]{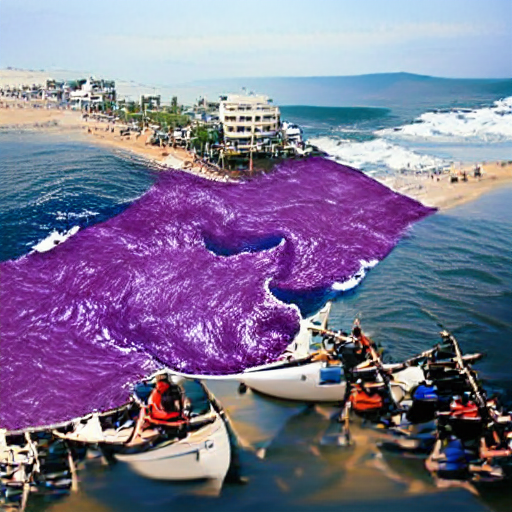}
\centering
\caption{Four images from the following prompts: Biophilic vertical gardens lining neighborhood roads, creating function and beautiful public spaces (left), Public spaces: solidarity-building. The intersection of oceans and relationships. Publicly accessible oceanic vistas  (center left), Holistic traditional medicine as an art form (center right), Dye the ocean purple to prevent global warming (right) }
\label{fig:images}
\end{figure*}

While all loosely related to the solarpunk theme, the written prompts from participants in the original activity had a remarkable amount of diversity. Some participants generated futuristic ideas, while others called for a return to traditional wisdom or practices. Four exemplary images, with their corresponding prompts are shown in Figure ~\ref{fig:images}. 
\subsubsection{User Interviews}
We conducted 10 interviews with individuals who participated in the original activity. We focused our interviews to participants who remembered the conversation they had from several months prior and still recalled the vision that they wrote down. 

We coded our 10 participant interviews for major themes distilled through repeated interviewing. In the process of synthesizing recurring themes across conversations, we noticed that many participants gleaned new insight as a result of viewing the image. We recorded when participants mentioned new information gleaned about their original idea after viewing the generated image. A majority of participants reported gaining new ideas as a result of viewing the image generated by their prompt, and this new information fell into two main categories.

The first was the emergence of new, unexpected ideas for implementing the vision they had written (40\% of interviews, n = 4). For example, the creator of the prompt ``Biophilic vertical gardens lining neighborhood roads, creating function and beautiful public spaces” already had a ``pretty concrete'' image of what their vision could look like given that biophilic community gardens already exist. However, they had imagined ``pumps and tubes'' as a visual feature given the complex engineering required to create such gardens. When viewing the generated image, the creator noted that the wall looked like it was made of a ``natural, rocky substrate'' rather than one with exposed hydroponic engineering (see Figure ~\ref{fig:images}, left). This lead the participant to note that perhaps the wall of the garden could in fact be beautiful and natural-looking as well as functional, and that this aesthetic would be an improvement over exposed pipes. 

The second main area of insight participants reported in interviews related to generating unexpected connections between a vision and lateral concepts (40\%, n = 4). For example, the prompt ``Public spaces: solidarity-building. The intersection of oceans and relationships. Publicly accessible oceanic vistas'' generated an image of humans on a beach (see Figure ~\ref{fig:images}, center left). Something like sand is present, but the lack of an actual ocean in the image was surprising to the prompt’s creator. However, the repeated visual motif of people in relationship yielded a new perspective for the creator, who then reflected on the importance of relationship for organizing in public space. This was not a framing he had not been considering as central to this prompt before being presented with the image. The lack of ocean in an image generated from a prompt with two references to the ocean could be considered low-fidelity and ultimately undesirable behavior from VQGAN+CLIP. Yet the unexpectedness of the image led the prompt's creator in a new direction that was ultimately valuable for ideation. As a result, we consider this a happy accident in the context of this exercise in social dreaming.

In another example, the participant who wrote ``Holistic traditional medicine as an art form'' reported noticing the centrality of hands as a visual motif in the image (see Figure  ~\ref{fig:images}, center right). Her interpretation was that the hands made the image ``focused on the making process'' and that the image ``emphasizes labor.'' She reported that the human labor aspect behind traditional medicine was not a major association she had with the topic before viewing the image.

Most (80\%, n = 8) interviewees mentioned at least one idea for a follow-up prompt to refine the image or clarify their vision. For example, the biophilic garden image featured a visual element on the bottom left that looked like a road. The prompt's creator noticed the curb and said they might use language like ``neighborhood path" instead of ``road" to generate an image with explicitly pedestrian streets in the future.

Despite the fact that it was easy for participants to identify the image created from their prompt (90\%, n = 9) out of a set of random images, most participants interpreted the generated images as being either partly or substantially different from what they imagined prior to seeing the image (70\%, n = 7). Three interviewees that did not gain any new insight from viewing the images were concerned that the image was too abstract to be useful. On the other hand, two other participants commented that they specifically appreciated the ``whimsical'' image of a technical concept.

\subsubsection{``When the place inspires the zine inspires the place''\protect\footnote{Quote by Sidebody \url{https://www.instagram.com/p/CbtJnWlloui/}}}
\begin{figure*}[h!]
\includegraphics[width=0.99\textwidth]{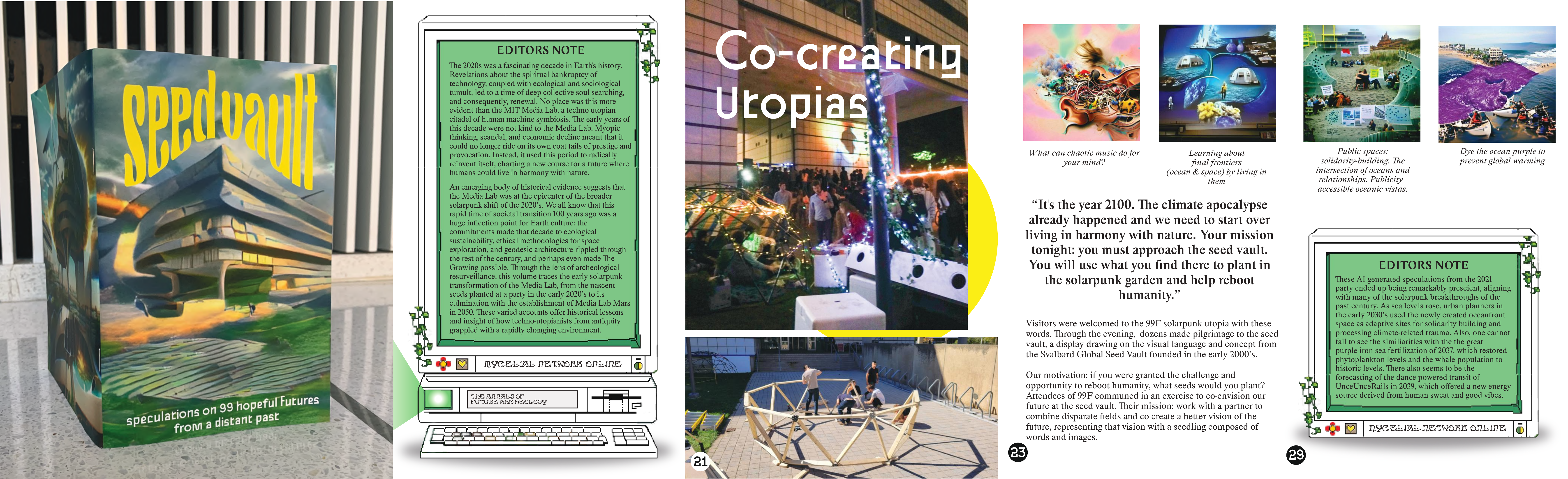}
\centering
\caption{Cover and four select pages from the Seedvault zine. Written in the style of speculative fiction, the zine takes place in the year 2121, recounting an event that took place 100 years ago and how the previous 100 years unfolded from there. }
\label{fig:zine}
\end{figure*}

In the wake of the solarpunk event, we created a zine outlining the project and showcasing all the co-written visions and corresponding GAN images to aid in community reflection (see Figure ~\ref{fig:zine}). We printed over 500 copies and distributed them to community stakeholders. As community members, we wanted to give these images back to the community, hoping they would act as a mirror for additional conversation and ``close the loop'' of the reflective process.

The event, co-creation experience, and zine all came at a time when our organization was emerging from the throes of the pandemic and actively setting its strategy and vision for the next decade and beyond. At such a critical inflection point, the combination of these three elements had several key cultural impacts. First, there was the physical aspect of people coming together to imagine and plant the seeds for the future after a long period of pandemic-induced isolation. Second, in an organization with varied interests and priorities, the process of brainstorming through images helped visualize and synthesize collective threads across the organization. Finally, cementing the images in familiar, tangible form of printed media provided a snapshot for both disseminating the outputs and orienting discussion around the organization's future \citep{perez2020towards}. The zine has since been used to communicate interests and priorities to both visitors and external stakeholders. 

\section{Discussion}
Initially, we predicted a main contribution of this work would be designing a tool to concretize visions that teams co-created. We found that occasionally the image did meaningfully crystallize a vision for the future in a literal way, and many participants noted the image gave them ideas for ways to implement their vision. However, we often found that it was the unexpected differences between the prompt and the generated image’s interpretation of it that yielded new insight for and excitement from participants. 
This suggests that it was the high-variance, low-fidelity behavior of VQGAN+CLIP that yielded these unexpected and whimsical differences, which in turn induced novel and creative insights. This finding has important implications for the design of image synthesis systems for computational creativity. As new systems become increasingly realistic (such as new models like DALL-E 2 \citep{ramesh2022hierarchical}), there is a danger that this increased fidelity will come at the cost of these unexpected quirks that we found actually sparked positive lateral thinking in our participants.

By offering a low-friction and provocative new way to integrate visual storytelling with community engagement, text-to-image models with collaboratively sourced prompts have an opportunity for cultural and organizational impact. This approach could be used in contexts as diverse as crafting organization mission statements, co-designing community interventions, and facilitating a mediation process. 

There are, however, dangers to such an approach. For one, AI-generated images are bounded by training data, which inherits historical biases and cultural practices \citep{ganimals, crawford2019excavating}. Therefore uncritically relying on model outputs as an ``oracle’’ may entrench users in existing inequities and stereotypes, rather than freeing them to envision radical new possibilities. Relatedly, if such an approach becomes commonplace, there is the risk that such visualization strategies could be a crutch if used too much, with users becoming overly reliant on a machine’s vision of the future rather than their own. Finally, there is the risk that anthropomorphizing the AI can undermine human credit and responsibility \citep{epstein2020gets}.

    It is also important to reflect on the notion of ``utopias,’’ which we used to frame this collaborative speculation. Literally meaning ``no place'' in the original Greek, utopias represent idealized non-existent or impossible societies. Rarely in utopian thinking are questions like ``A utopia for whom?'' or ``A utopia at what cost?'' considered, which in turn leads to imagined futures that are culturally homogeneous and can perpetuate existing inequities. Furthermore, the very concept of utopias imagines a future society different from and evolved upon the current one, a notion rooted in endless growth, a potentially colonial and capitalistic value \citep{morrison2017decolonizing}. Rather than envisioning radical new alternatives to be manifested, many indigenous cultures instead hold a worldview that humans are a part of a complex web of ecological relationships that can exist an a perpetual steady-state equilibrium \citep{kimmerer2013braiding, sepie2017more}. In the prompts we received,  we saw varied interpretations of and priorities for utopias: some participants longed for a return to local economies or traditional medicine, while others dreamt of futuristic space suits and wind turbines. Finally, it is important to note that utopias are but one frame for imaginative idea visualization. While we used the idea of utopia to foster brainstorming about the future, we believe the paradigm could instead focus inward for any community or space looking to collaboratively speculate about its future.

Our work has several limitations which open up exciting possibilities for future work. For one, we focused on one particular community -- the MIT Media Lab, a technology-focused research institution embedded in a university.  Future work could explore how this approach works for other communities with distinct cultures and practices, and how the effectiveness of the approach varies across community contexts.  In addition, we relied on analog media throughout the process: from handwritten prompts by participants and a paper zine distributed in person to us showing participants generated images asynchronously in the evaluation phase. Future work could explore digital versions of these methods, such as an online interface for collaborative prompt generation, an online gallery for the prompts and their corresponding images, or real-time dialogue with a generative model. We also did not include in-depth explanation for how the generative model worked. Future work should consider more explicitly users' understanding of the involved technology, as well as consider other types of models, such as physically-informed models of climate futures \citep{lutjens2021physically, lutjens2021pce} or those that integrate across larger, more complex systems \citep{lavin2021technology}. Finally, there is the possibility of using this method in diverse settings. We hope that this approach could be applied to other community contexts, whether it be designs for a local community garden or bold new tactics to fight the global climate emergency.

\section*{Acknowledgements}
We thank Eyal Perry, Micah Epstein, Amy Smith, Simon Colton, Rubez Chong, Océane Boulais, Maggie Hughes, Valdemar Danry and Tobin South for helpful feedback, comments and support. We thank all participants for their time.
\section*{Author Contributions}
ZE, HS, and DN conceptualized the project. HS conducted the user interviews. ZE and HS wrote the paper, with input from DN.  






\bibliographystyle{iccc}
\bibliography{iccc}

\begin{thebibliography}{}

\bibitem[\protect\citeauthoryear{Berns and Colton}{2020}]{berns2020bridging}
Berns, S., and Colton, S.
\newblock 2020.
\newblock Bridging generative deep learning and computational creativity.
\newblock In {\em ICCC},  406--409.

\bibitem[\protect\citeauthoryear{Colton \bgroup et al.\egroup
  }{2021}]{colton2021generative}
Colton, S.; Smith, A.; Berns, S.; Murdock, R.; and Cook, M.
\newblock 2021.
\newblock Generative search engines: Initial experiments.
\newblock In {\em ICCC}.

\bibitem[\protect\citeauthoryear{Compton and Mateas}{2015}]{compton2015casual}
Compton, K., and Mateas, M.
\newblock 2015.
\newblock Casual creators.
\newblock In {\em ICCC},  228--235.

\bibitem[\protect\citeauthoryear{Crawford and
  Paglen}{2019}]{crawford2019excavating}
Crawford, K., and Paglen, T.
\newblock 2019.
\newblock Excavating ai: The politics of images in machine learning training
  sets.
\newblock {\em AI and Society}.

\bibitem[\protect\citeauthoryear{Dunne and Raby}{2013}]{dunne2013speculative}
Dunne, A., and Raby, F.
\newblock 2013.
\newblock {\em Speculative everything: design, fiction, and social dreaming}.
\newblock MIT press.

\bibitem[\protect\citeauthoryear{Epstein \bgroup et al.\egroup
  }{2020a}]{epstein2020interpolating}
Epstein, Z.; Boulais, O.; Gordon, S.; and Groh, M.
\newblock 2020a.
\newblock Interpolating gans to scaffold autotelic creativity.
\newblock {\em International Conference on Computational Creativity Causal
  Creator Workshop}.

\bibitem[\protect\citeauthoryear{Epstein \bgroup et al.\egroup
  }{2020b}]{epstein2020gets}
Epstein, Z.; Levine, S.; Rand, D.~G.; and Rahwan, I.
\newblock 2020b.
\newblock Who gets credit for ai-generated art?
\newblock {\em Iscience} 23(9):101515.

\bibitem[\protect\citeauthoryear{{Ganimals Blog}}{2020}]{ganimals}
{Ganimals Blog}.
\newblock 2020.
\newblock Beware the training data: The barracuda effect.
\newblock ganimals.media.mit.edu.

\bibitem[\protect\citeauthoryear{Kimmerer}{2013}]{kimmerer2013braiding}
Kimmerer, R.
\newblock 2013.
\newblock {\em Braiding sweetgrass: Indigenous wisdom, scientific knowledge and
  the teachings of plants}.
\newblock Milkweed editions.

\bibitem[\protect\citeauthoryear{Lavin \bgroup et al.\egroup
  }{2021}]{lavin2021technology}
Lavin, A.; Gilligan-Lee, C.~M.; Visnjic, A.; Ganju, S.; Newman, D.; Ganguly,
  S.; Lange, D.; Baydin, A.~G.; Sharma, A.; Gibson, A.; et~al.
\newblock 2021.
\newblock Technology readiness levels for machine learning systems.
\newblock {\em arXiv preprint arXiv:2101.03989}.

\bibitem[\protect\citeauthoryear{L{\"u}tjens \bgroup et al.\egroup
  }{2021a}]{lutjens2021pce}
L{\"u}tjens, B.; Crawford, C.~H.; Veillette, M.; and Newman, D.
\newblock 2021a.
\newblock Pce-pinns: Physics-informed neural networks for uncertainty
  propagation in ocean modeling.
\newblock {\em arXiv preprint arXiv:2105.02939}.

\bibitem[\protect\citeauthoryear{L{\"u}tjens \bgroup et al.\egroup
  }{2021b}]{lutjens2021physically}
L{\"u}tjens, B.; Leshchinskiy, B.; Requena-Mesa, C.; Chishtie, F.;
  D{\'\i}az-Rodr{\'\i}guez, N.; Boulais, O.; Sankaranarayanan, A.; Pi{\~n}a,
  A.; Gal, Y.; Ra{\"\i}ssi, C.; et~al.
\newblock 2021b.
\newblock Physically-consistent generative adversarial networks for coastal
  flood visualization.
\newblock {\em arXiv preprint arXiv:2104.04785}.

\bibitem[\protect\citeauthoryear{Morrison}{2017}]{morrison2017decolonizing}
Morrison, M.~I.
\newblock 2017.
\newblock {\em Decolonizing Utopia: Indigenous Knowledge and Dystopian
  Speculative Fiction}.
\newblock University of California, Riverside.

\bibitem[\protect\citeauthoryear{P{\'e}rez~y P{\'e}rez and
  Ackerman}{2020}]{perez2020towards}
P{\'e}rez~y P{\'e}rez, R., and Ackerman, M.
\newblock 2020.
\newblock Towards a methodology for field work in computational creativity.
\newblock {\em New Generation Computing} 38(4):713--737.

\bibitem[\protect\citeauthoryear{Radford \bgroup et al.\egroup
  }{2021}]{radford2021learning}
Radford, A.; Kim, J.~W.; Hallacy, C.; Ramesh, A.; Goh, G.; Agarwal, S.; Sastry,
  G.; Askell, A.; Mishkin, P.; Clark, J.; et~al.
\newblock 2021.
\newblock Learning transferable visual models from natural language
  supervision.
\newblock In {\em International Conference on Machine Learning},  8748--8763.
\newblock PMLR.

\bibitem[\protect\citeauthoryear{Rafner \bgroup et al.\egroup
  }{2020}]{rafner2020crea}
Rafner, J.; Hjorth, A.; Risi, S.; Philipsen, L.; Dumas, C.; Biskj{\ae}r, M.~M.;
  Noy, L.; Tyl{\'e}n, K.; Bergenholtz, C.; Lynch, J.; et~al.
\newblock 2020.
\newblock crea. blender: A neural network-based image generation game to assess
  creativity.
\newblock In {\em Extended Abstracts of the 2020 Annual Symposium on
  Computer-Human Interaction in Play},  340--344.

\bibitem[\protect\citeauthoryear{Rafner \bgroup et al.\egroup
  }{2021}]{rafner2021utopian}
Rafner, J.; Langsford, S.; Hjorth, A.; Gajdacz, M.; Philipsen, L.; Risi, S.;
  Simon, J.; and Sherson, J.
\newblock 2021.
\newblock Utopian or dystopian?: using a ml-assisted image generation game to
  empower the general public to envision the future.
\newblock In {\em Creativity and Cognition},  1--5.

\bibitem[\protect\citeauthoryear{Ramesh \bgroup et al.\egroup
  }{2022}]{ramesh2022hierarchical}
Ramesh, A.; Dhariwal, P.; Nichol, A.; Chu, C.; and Chen, M.
\newblock 2022.
\newblock Hierarchical text-conditional image generation with clip latents.
\newblock {\em arXiv preprint arXiv:2204.06125}.

\bibitem[\protect\citeauthoryear{Salda{\~n}a}{2013}]{saldana2021coding}
Salda{\~n}a, J.
\newblock 2013.
\newblock {\em The coding manual for qualitative researchers}.
\newblock SAGE.

\bibitem[\protect\citeauthoryear{Sepie}{2017}]{sepie2017more}
Sepie, A.~J.
\newblock 2017.
\newblock More than stories, more than myths: Animal/human/nature (s) in
  traditional ecological worldviews.
\newblock {\em Humanities} 6(4):78.

\end{thebibliography}

\end{document}